\documentclass[draft]{agujournal2019}
\usepackage{url} 
\usepackage[finalnew]{trackchanges} 

\usepackage{soul}
\draftfalse

\journalname{JGR: Oceans}

\begin{document}

%
%


\title{Comment on ``The wave-driven current in coastal canopies" by M. Abdolahpour et al.}

%
%




\authors{Mitul Luhar\affil{1}}


\affiliation{1}{Department of Aerospace and Mechanical Engineering, University of Southern California, Los Angeles, CA 90089}




\correspondingauthor{Mitul Luhar}{luhar@usc.edu}




\begin{keypoints}
\item Prior studies show the emergence of a strong wave-driven mean current in submerged canopies.
\item Two different driving mechanisms (and predictive formulations) have been proposed for this mean flow.
\item This comment provides a modified model and scaling arguments to reconcile these formulations.
\end{keypoints}

%
%

%
%


\begin{abstract}
Laboratory and field measurements made over the past decade have shown the presence of a strong wave-driven mean current in submerged vegetation canopies. \citeA{luhar2010wave} suggested that this mean current is analogous to the streaming flow generated in wave boundary layers over bare beds, and developed a simple energy and momentum balance model to predict its magnitude. However, this model predicts that the magnitude of the mean current does not depend on canopy spatial density, which is inconsistent with the measurements made by \citeA{abdolahpour2017wave} in recent laboratory experiments. Motivated by observations that the wave-driven mean flow is most pronounced at the canopy interface, \citeA{abdolahpour2017wave} proposed an alternate explanation for its origin: that it is driven by the vertical heterogeneity in orbital motion created by canopy drag. Such heterogeneity can give rise to incomplete particle orbits near the canopy interface and a Lagrangian mean current analogous to Stokes drift in the direction of wave propagation. A model guided by this physical insight and dimensional analysis is able to generate much more accurate predictions. This comment aims to reconcile these two different models for the wave-driven mean flow in submerged canopies.
\end{abstract}

\section*{Plain Language Summary}
Previous laboratory and field measurements have shown the emergence of a strong wave-driven mean current in submerged canopies of aquatic vegetation. By controlling the rate of water renewal in coastal canopies, this mean current could play a vital role in mediating the valuable ecosystem services provided by vegetated systems such as seagrass beds (e.g., nutrient and carbon uptake). However, two different driving mechanisms have been identified in previous studies for this wave-induced mean current, which has led to two distinct predictive formulations for its magnitude. This brief contribution describes a modified model that aims to reconcile these formulations and, thereby, clarify the hydrodynamic mechanism driving the mean current.

\section{Background}
Many of the ecosystem services provided by aquatic vegetation (e.g., nutrient and carbon uptake, oxygen production) are limited by the rate of water exchange between the canopy and the surrounding environment. The wave-driven mean current discussed in \citeA{abdolahpour2017wave} could be an important mechanism for such water renewal in submerged coastal canopies. The goal of this comment is to elaborate on---and reconcile---the different models proposed by \citeA{luhar2010wave} and \citeA{abdolahpour2017wave} to describe this wave-driven mean flow. 

It is well established that, in the absence of canopies, progressive surface waves can give rise to mean currents through two different mechanisms: Stokes drift and boundary-layer streaming. Stokes drift is the net flow that results from incomplete particle orbits near the water surface in the presence of waves \cite{van2017stokes}. Boundary layer streaming is driven by a wave stress arising from nonzero temporal correlation between the horizontal and vertical components of the oscillatory flow \cite{longuet1953mass,scandura2007steady}. In addition to the different driving mechanisms, there is another important distinction between these phenomena. While boundary layer streaming can be measured using fixed-point instruments, Stokes drift is a Lagrangian phenomenon that is difficult to capture in such Eulerian measurements \cite{umeyama2012eulerian}. Indeed, Stokes drift is often defined more generally as the difference between the average Lagrangian velocity of a fluid parcel and the mean Eulerian velocity of the fluid. 

\citeA{luhar2010wave} suggest that the wave-driven mean current observed in submerged canopies is analogous to boundary layer streaming, i.e., driven by a wave stress. However, \citeA{abdolahpour2017wave} suggest that it is analogous to Stokes drift, i.e., arising from incomplete particle orbits near the canopy interface. These differing interpretations have led to two distinct formulations that predict the magnitude of the mean current. These two formulations are summarized and discussed briefly below. A modified model that aims to reconcile both formulations is presented in the next section.

\citeA{luhar2010wave} used the following energy and momentum balance arguments to predict the magnitude of the wave-driven mean flow. The time-averaged wave stress driving the mean current was estimated based on two key assumptions. First, that the wave energy dissipated by vegetation drag inside the canopy is balanced by the net work done by pressure at the canopy interface, $-\overline{p_w w_w}$. Second, that the relationship between the pressure and horizontal velocity above the canopy is described adequately by linear wave theory, $p_w = \rho(\omega/k)u_w$. Here, $p_w$, $u_w$, and $w_w$ are the wave-induced pressure, horizontal velocity, and vertical velocity fields; $\rho$ is the fluid density; $\omega$ is the wave frequency; and $k$ is the wavenumber. An overbar denotes a time average.  With these assumptions, the wave stress was estimated to be:
\begin{equation}\label{eq:wavestress}
    \tau_w = -\rho\overline{u_w w_w} \approx  \frac{k}{\omega} \frac{2}{3\pi} \rho \int_{0}^{h_v} C_{Dw} a U_c^3 dz,
\end{equation}
in which $h_v$ is the height of the region occupied by the plants, $C_{Dw}$ is a wave drag coefficient for the vegetation, $a$ is the vegetation frontal area per unit volume, $U_c$ is the amplitude of the horizontal oscillatory velocity inside the canopy, and $z$ is the coordinate normal to the bed. Next, the momentum transferred into the canopy by this wave stress was assumed to be balanced by the mean vegetation drag induced by the wave-driven current,
\begin{equation}\label{eq:dragbalance}
  \tau_w \approx \frac{1}{2} \rho \int_{0}^{h_v} C_{Dc} a \overline{u}_c^2 dz,
\end{equation}
where $C_{Dc}$ is a current drag coefficient and $\overline{u}_c$ is the time-averaged mean flow inside the canopy. Finally, assuming that the quantities $h_v$, $C_{Dc}$, $C_{Dw}$, $a$, $U_c$, and $\overline{u}_c$ are approximately constant over the height of the canopy, equations~(\ref{eq:wavestress}) and (\ref{eq:dragbalance}) were combined and simplified to yield the following expression for the magnitude of the mean current:
\begin{equation}\label{eq:luhar_uc}
    \overline{u}_c = \sqrt{\frac{4}{3\pi}\frac{C_{Dw}}{C_{Dc}}\frac{k}{\omega}U_c^3}.
\end{equation}
Clearly, this expression involves a number of assumptions and simplifications. For an extended discussion of these issues, the reader is referred to \citeA{luhar2010wave}.

Equation~(\ref{eq:luhar_uc}), with $C_{Dw}/C_{Dc}$ set to 1 for simplicity, was shown to generate reasonable predictions for the wave-driven currents observed by \citeA{luhar2010wave} in laboratory experiments over model seagrass canopies. However, it fails to yield accurate predictions for the mean currents that have been measured in subsequent field studies over seagrass beds \cite{luhar2013field} and laboratory studies involving rigid and flexible model vegetation \cite{abdolahpour2017wave}. In particular, the formulation developed by \citeA{luhar2010wave} is inconsistent with the laboratory measurements made by \citeA{abdolahpour2017wave} in two important ways. First, it predicts that the magnitude of the mean current does not depend on canopy density. Second, the model assumes that the mean current is distributed over the entire canopy height. Measurements made by \citeA{abdolahpour2017wave} show that for rigid model vegetation the mean current is confined to the canopy interface, with a vertical extent that is comparable to the vertical orbital excursion, $\xi_T$.  For flexible model vegetation, the mean current is most pronounced at an elevation corresponding roughly to the height of the canopy in its most pronated state over a wave-cycle.  Moreover, for both rigid and flexible canopies, the magnitude of the mean current increases as the canopy frontal area parameter, $a$, increases (or equivalently, as the canopy drag length scale $L_D \approx a^{-1}$ decreases; see equation (5) in \citeA{abdolahpour2017wave}).

Motivated by the observation that the wave-induced mean flow is most pronounced at the canopy interface, \citeA{abdolahpour2017wave} proposed an alternate interpretation for its origin: that it is driven by the vertical heterogeneity in orbital motion created by canopy drag. Since drag reduces orbital velocities within the canopy, \citeA{abdolahpour2017wave} argued that a fluid particle near the interface would experience higher shoreward velocity under the wave crest and reduced offshore velocities under the wave trough (see figure 3 in \citeA{abdolahpour2017wave}). This would result in open particle orbits and a Lagrangian mean current in the direction of wave propagation, similar to Stokes drift \cite{jacobsen2016wave}. This interpretation was further supported by the observation that maximum mean velocities measured by \citeA{abdolahpour2017wave} scaled with the difference in orbital velocities above and below the interface. By combining this physical insight with dimensional reasoning and fits to laboratory measurements, \citeA{abdolahpour2017wave} developed the following expression to predict the maximum mean current:
\begin{equation}\label{eq:abdolahpour_umax}
    \overline{u}_{max} = 0.5 U_\infty^{rms} \left(\frac{\xi_T}{L_D}\right)^{0.3}.
\end{equation}
Here, $U_\infty^{rms}$ is the root-mean-square value of the horizontal orbital velocity at the canopy interface. Note that equation~(3) makes use of the \textit{in-canopy} horizontal orbital velocity, $U_c$, while equation~(4) depends on the velocity at the canopy interface, $U_\infty$. \citeA{abdolahpour2017wave} show that equation~(\ref{eq:abdolahpour_umax}) yields much more accurate predictions for the mean currents measured in prior laboratory and field experiments compared to the expression shown in equation~(\ref{eq:luhar_uc}). Further, equation~(\ref{eq:abdolahpour_umax}) explicitly accounts for canopy density through the drag length scale $L_D$.

Equation~(\ref{eq:abdolahpour_umax}) is a clear step forward in terms of predictive capability. However, the Lagrangian interpretation proposed by \citeA{abdolahpour2017wave} remains problematic. This is because all prior measurements for the wave-driven mean current have come from fixed-point Acoustic Doppler Velocimeters. In other words, the mean current is clearly observed in Eulerian measurements. It is generally accepted that any mean currents resulting from a Stokes drift-like phenomenon appear as the difference between the mean Lagrangian and Eulerian velocities. So, if the mean current was analogous to Stokes drift in origin (i.e., driven by spatial heterogeneity in orbital motion), measuring it using fixed-point instruments would be challenging. As demonstrated in \citeA{umeyama2012eulerian}, this would require Particle Tracking Velocimetry techniques, or spatial interpolation and Lagrangian integration of velocity fields from Particle Image Velocimetry. 

Thus, the streaming flow interpretation proposed by \citeA{luhar2010wave} is supported by the fact that all prior measurements of the wave-driven current in submerged canopies have come from fixed-point instruments. However, the expression shown in equation~(\ref{eq:abdolahpour_umax}), developed by \citeA{abdolahpour2017wave} using Lagrangian arguments, generates significantly better predictions. A modified model and scaling arguments that can potentially reconcile this discrepancy are presented next. 

\section{Modified Model}
As noted earlier, the expression shown in equation (\ref{eq:luhar_uc}), developed by \citeA{luhar2010wave} using energy and momentum balance arguments, has two important deficiencies. First, it predicts that the magnitude of the mean current is not dependent on canopy density. Second, it assumes that the streaming flow is distributed across the entire height of the canopy. Since the experimental observations of \citeA{abdolahpour2017wave} indicate that the wave-driven mean current is most pronounced at the canopy interface, the distributed drag formulation shown on the right-hand side of equation~(\ref{eq:dragbalance}) can arguably be replaced with an interfacial friction formulation dependent on the maximum mean current, i.e., 
\begin{equation}\label{eq:friction}
  \tau_w \approx \frac{1}{2} \rho C_{f} \overline{u}_{max}^2, 
\end{equation}
in which $C_f$ is a coefficient representative of the frictional resistance in the upper region of the canopy. An alternative interpretation consistent with prior work on submerged canopy flows \cite{nepf2012flow} would be that the mean current penetrates into the canopy to a vertical distance comparable to the drag length scale $L_D$, such that the integral in equation~(\ref{eq:dragbalance}) scales as $\int_{0}^{h_v} C_{Dc} a \overline{u}_c^2 dz \sim C_{Dc} a L_D \overline{u}_{max}^2 \approx C_{Dc} \overline{u}_{max}^2$. This argument yields an expression similar to that shown in equation~(\ref{eq:friction}). 

Combining equation~(\ref{eq:wavestress}) with the modified formulation shown in equation~(\ref{eq:friction}), and again assuming that the quantities $C_{Dw}$, $a$, and $U_c$ are uniform over the height of the canopy leads to:
\begin{equation}
    \frac{k}{\omega} \frac{2}{3\pi} \rho C_{Dw} a h_v U_c^3 \approx \frac{1}{2} \rho C_{f} \overline{u}_{max}^2,
\end{equation}
or equivalently
\begin{equation}\label{eq:newdragbalance}
     \overline{u}_{max} \approx U_c \sqrt{\frac{4}{3\pi}\frac{k}{\omega}\frac{C_{Dw}}{C_f} a h_v U_c}.
\end{equation}
Thus, accounting for the localization of the mean current in the upper region of the canopy also introduces a density dependence in the predicted magnitude. More specifically, equation~(\ref{eq:newdragbalance}) predicts that the magnitude of the mean current increases as the vegetation frontal area parameter, $a$, increases, which is broadly consistent with the experimental observations of \citeA{abdolahpour2017wave}. Unlike equations~(3)-(4), equation~(7) predicts that the magnitude of the mean current also depends on the canopy height, $h_v$.  This parameter can be difficult to define for real aquatic canopies that are flexible and move in response to the fluid flow. For flexible vegetation, the effective blade length concept used in recent studies could be a useful surrogate for $h_v$ \cite{luhar2016wave,luhar2017seagrass,lei2019blade}.

\begin{figure}[t]
    \centering
    \includegraphics[width=12cm]{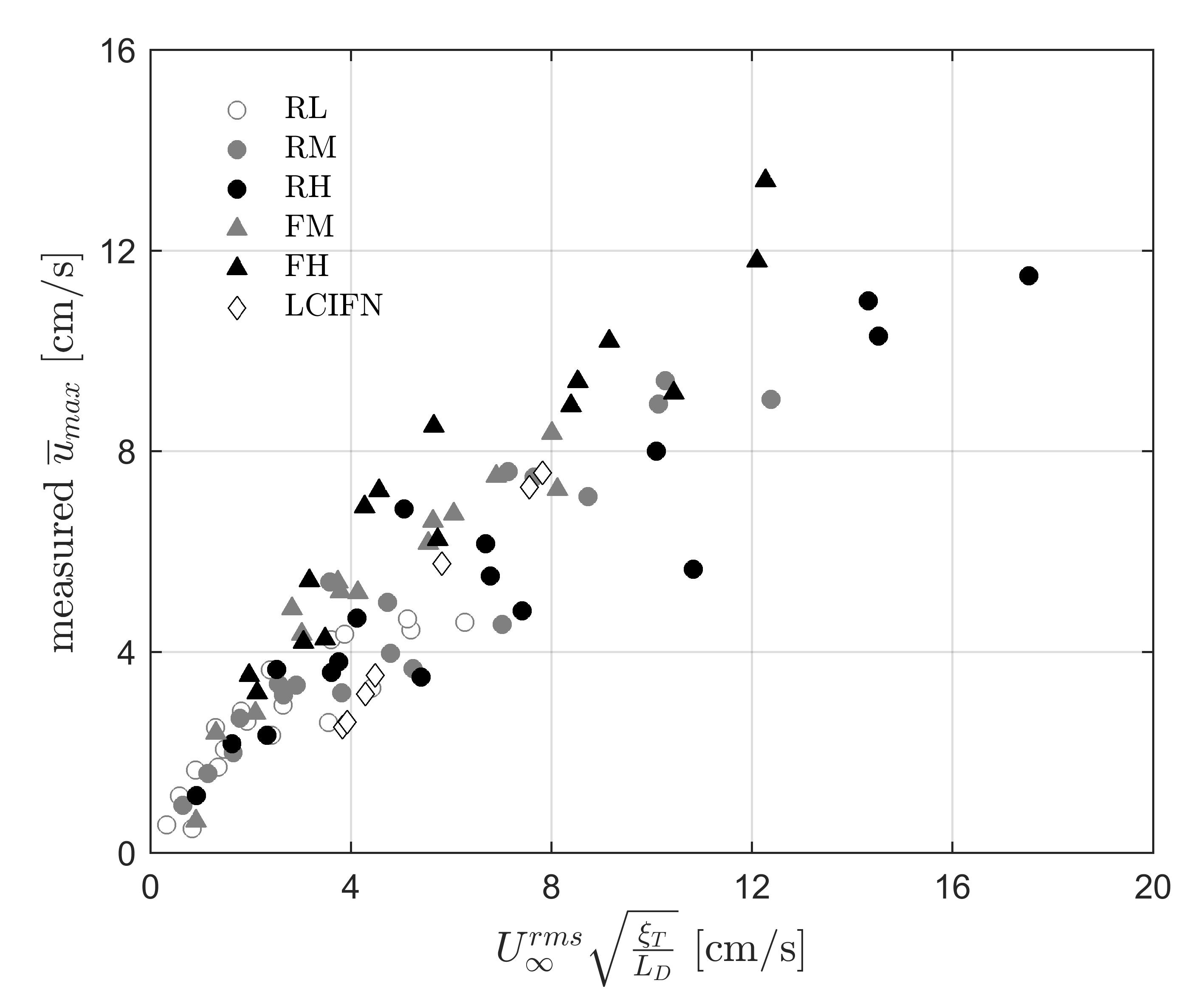}
    \caption{Comparison between the physically-motivated scaling shown in equation~(\ref{eq:newumax}) and measured maximum mean currents.  The datasets RL, RM, and RH correspond to the low, medium, and high-density rigid vegetation measurements from \citeA{abdolahpour2017wave}; FM and FH correspond to the medium and high-density flexible vegetation tests. Data from the experiments of \citeA{luhar2010wave} are also shown (abbreviated as LCIFN).  \add{The amplitude of the vertical orbital excursion is estimated as $\xi_T \approx k h_v U_\infty/\omega$ and the drag length scale is estimated as $L_D \approx a^{-1}$.}}\label{fig:ucmax}
\end{figure}

Importantly, it can be shown that the modified formulation in equation~(\ref{eq:newdragbalance}) is similar in form to the expression proposed by \citeA{abdolahpour2017wave}. 
Previous work shows that wave-induced oscillatory flows are not damped significantly inside vegetation canopies if the horizontal orbital excursion is smaller than, or comparable to, the drag length scale, $L_D$ (see figure 4 in \citeA{lowe2005oscillatory} and tables 1 and 3 in \citeA{luhar2010wave}).  For such conditions, the magnitude of the in-canopy orbital velocity is expected to be similar to that at the interface, $U_c \approx U_\infty$. Further, for cases in which canopy height is much smaller than the wavelength, the vertical orbital excursion at the interface can be approximated as $\xi_T = W_\infty/\omega \approx k h_v U_\infty / \omega$. Here, $W_\infty \approx k h_v U_\infty$ is the vertical orbital velocity at the canopy interface. With these factors in mind, and noting that $a \approx L_D^{-1}$, the expression in equation~(\ref{eq:newdragbalance}) yields the following physically-motivated scaling for the maximum mean current:
\begin{equation}\label{eq:newumax}
    \overline{u}_{max} \sim U_c \sqrt{a \frac{k h_v}{\omega} U_c} \sim U_\infty^{rms} \left(\frac{\xi_T}{L_D}\right)^{0.5}.
\end{equation}
This expression bears a strong resemblance to the empirical formulation in equation~(\ref{eq:abdolahpour_umax}) developed by \citeA{abdolahpour2017wave}. The exponent for the $(\xi_T/L_D$) term is different, though this difference can perhaps be attributed to variations in the wave drag and friction coefficients, $C_{Dw}$ and $C_f$, with flow conditions. Figure~\ref{fig:ucmax} confirms that the scaling shown in equation~(\ref{eq:newumax}) describes the maximum mean currents measured by \citeA{luhar2010wave} and \citeA{abdolahpour2017wave} reasonably well. The best-fit line obtained via linear regression ($ \overline{u}_{max} = 0.9 U_\infty^{rms} \sqrt{\xi_T/L_D}$) yields good agreement with the measured data ($r^2 = 0.74$). A power law of the form shown in equation~(\ref{eq:abdolahpour_umax}) leads to slightly better agreement with the measurements ($r^2 = 0.83$), albeit with two fitted parameters.

Thus, the predictive success of the formulation proposed by \citeA{abdolahpour2017wave} does not require the wave-driven mean current in submerged canopies to be Lagrangian in origin. A similar expression can also be developed using momentum and energy balance arguments similar to those made by \citeA{luhar2010wave}. Therefore, given that all prior measurements of the wave-driven mean flow in submerged canopies have come from fixed-point measurements, the (Eulerian) streaming flow interpretation proposed by \citeA{luhar2010wave} remains more appropriate. This streaming flow interpretation is also supported to some extent by recent numerical simulations that reproduce the emergence of a wave-driven mean flow in submerged canopies \cite{chen2019eulerian,chen2019wave}. Specifically, \citeA{chen2019wave} show that the Lagrangian mean velocity estimated at the canopy interface via particle tracking in the numerical simulations is approximately equal to the Eulerian mean velocity, i.e., Stokes drift is relatively small. It must be emphasized that this comment does not preclude the presence of the Stokes drift-like mean current described by \citeA{abdolahpour2017wave} in wave-driven flows over submerged canopies. Indeed, previous analytical efforts support its existence \cite{jacobsen2016wave}. However, the wave-driven mean currents measured in previous studies are unlikely to be a manifestation of this Lagrangian phenomenon. Perhaps future numerical simulations and experiments that involve explicit wave stress measurements or Lagrangian tracking of fluid parcels will provide greater insight into the relative importance of both mechanisms in driving mean flows over submerged canopies.

\acknowledgments
No new data were generated for this contribution. The data used in Figure~\ref{fig:ucmax} were sourced from Tables 1 and 2 in \citeA{abdolahpour2017wave} and from Table 1 in \citeA{luhar2010wave}.


%
%

\bibliography{references}

%
%
%
%
%

\end{document}